\documentclass{aa}  

\usepackage{graphicx}
\usepackage{color}
\usepackage{amssymb}
\usepackage[varg]{txfonts}
\begin{document}

%%%%%%%%%%%%%%%%%%
%%%%%%%%%%%%%%%%%%
%%%Title and authors
%%%%%%%%%%%%%%%%%%
%%%%%%%%%%%%%%%%%%
\title{Detection of volatiles undergoing sublimation from 67P/Churyumov-Gerasimenko coma particles using ROSINA/COPS}
\subtitle{II. The nude gauge}

\author{B. Pestoni\inst{1} \and K. Altwegg\inst{1} \and H. Balsiger\inst{1}\thanks{Deceased 19 January 2021.} \and N. H{\"a}nni\inst{1} \and M. Rubin\inst{1} \and I. Schroeder\inst{1} \and M. Schuhmann\inst{1} \and S. Wampfler\inst{2}}

\institute{Physics Institute, Space Research \& Planetary Sciences, University of Bern, Sidlerstrasse 5, 3012 Bern, Switzerland\\ \email{boris.pestoni@space.unibe.ch}
\and
Center for Space and Habitability, University of Bern, Gesellschaftsstrasse 6, 3012 Bern, Switzerland}

\date{Received XXX / Accepted YYY}

%%%%%%%%%%%%%%%%%%
%%%%%%%%%%%%%%%%%%
%%%Abstract
%%%%%%%%%%%%%%%%%%
%%%%%%%%%%%%%%%%%%
 
\abstract
{In an earlier study, we reported that the ram gauge of the COmet Pressure Sensor (COPS), one of the three instruments of the Rosetta Orbiter Spectrometer for Ion and Neutral Analysis (ROSINA), could be used to obtain information about the sublimating content of icy particles, made up of volatiles and conceivably refractories coming from comet 67P/Churyumov-Gerasimenko.}
{In this work, we extend the investigation to the second COPS gauge, the nude gauge. 
In particular, we analyse the volume of the volatile content of coma particles, along with a search for possible dependencies between the nude gauge detection rate (i.e. the rate at which icy particles are detected by the nude gauge) and the position of the Rosetta spacecraft. We also investigate the correlations of the nude gauge detection rate with the quantities associated with cometary activity.}
{We inspected the density measurements made by the nude gauge for features attributable to the presence of icy particles inside the instrument. These data were then analysed statistically based on the amplitude of the feature and on the position of the spacecraft at the time of detection.}
{Although it was not originally designed for such a purpose, the COPS nude gauge has been able to detect $\sim$67000 features generated by the sublimation of the volatile content of icy particles. The nude gauge detection rate follows a trend that is inversely proportional to the heliocentric distance. This result is interpreted as a confirmation of a possible relation between the nude gauge detection rate and cometary activity. Thus, we compared the former with parameters related to cometary activity and obtained significant correlations, indicating that the frequency of icy particle detection is driven by cometary activity. Furthermore, by representing the volatile part of the icy particles as equivalent spheres with a density of 1 g cm$^{-3}$, we obtained a range of diameters between 60 and 793 nanometres, with the smaller ones ($<390\,\mathrm{nm}$ in diameter) having a size distribution power index of $-4.79\pm 0.26$.}
{}

%%%%%%%%%%%%%%%%%%
%%%%%%%%%%%%%%%%%%
%%%Keywords
%%%%%%%%%%%%%%%%%%
%%%%%%%%%%%%%%%%%%

\titlerunning{Detection of volatiles from 67P coma particles using COPS-NG}
\maketitle

%%%%%%%%%%%%%%%%%%
%%%%%%%%%%%%%%%%%%
%%%Introduction
%%%%%%%%%%%%%%%%%%
%%%%%%%%%%%%%%%%%%

\section{Introduction}\label{sec:introduction}
Although the Rosetta mission was concluded in 2016, the vast quantities of data collected by its 21 instruments \citep[][]{Vallat_et_al_2017,Boehnhardt_et_al_2017} have enabled the continuation of studies concerning the Jupiter-family comet 67P/Churyumov-Gerasimenko (hereafter, 67P).

In \citet[hereafter, Part I]{Pestoni_et_al_2021}, we show that the ram gauge (hereafter, RG) of the COmet Pressure Sensor (COPS), while not originally designed for this purpose, was able to detect the volatile content of icy particles. COPS is one of three sensors in the Rosetta Orbiter Spectrometer for Ion and Neutral Analysis instrument package \citep[ROSINA,][]{Balsiger_et_al_2007}. Icy particles are defined as dust particles containing a condensed volatile component and a refractory component. The distinction between the two components is their different condensation point. When the volatile content of icy particles becomes sublimated upon entering the RG, there is a sudden increase in the density measured by the instrument. By analysing its subsequent decrease back to a nominal RG signal -- which is the nominal density background, generated by gaseous coma species, that is observed by the RG -- we were able to establish that there are three types of icy particles. This can be understood as different volatile compositions of the icy particles, with possible further complications due to
their differing morphologies. Assuming that the volatiles are composed solely of water (density of 1 g cm$^{-3}$), the sizes of the volatile content of the icy particles would be on the order of hundreds of nanometres in magnitude.

The results obtained from the RG measurements are complementary to those from other Rosetta instruments dedicated to the investigation of dust, such as the Micro-Imaging Dust Analysis System \citep[MIDAS,][]{Riedler_2007}, the Grain Impact Analyzer and Dust Accumulator \citep[GIADA,][]{Colangeli_2007}, and the COmetary Secondary Ion Mass Analyser \citep[COSIMA,][]{kissel_2007}. However, the RG is unable to measure the refractory component. Therefore, we must refer to the other aforementioned instruments for details on the refractories; see discussions in \citet{Rotundi_et_al_2015,Fulle_2015,bardyn_2017,Mannel_et_al_2019}. Nevertheless, COPS provides an additional opportunity for investigating particles of cometary origin. The classification scheme of dust particles proposed by \citet{Guttler_et_al_2019} is applied throughout this work.

\begin{figure}
\centering
\includegraphics[width=\hsize]{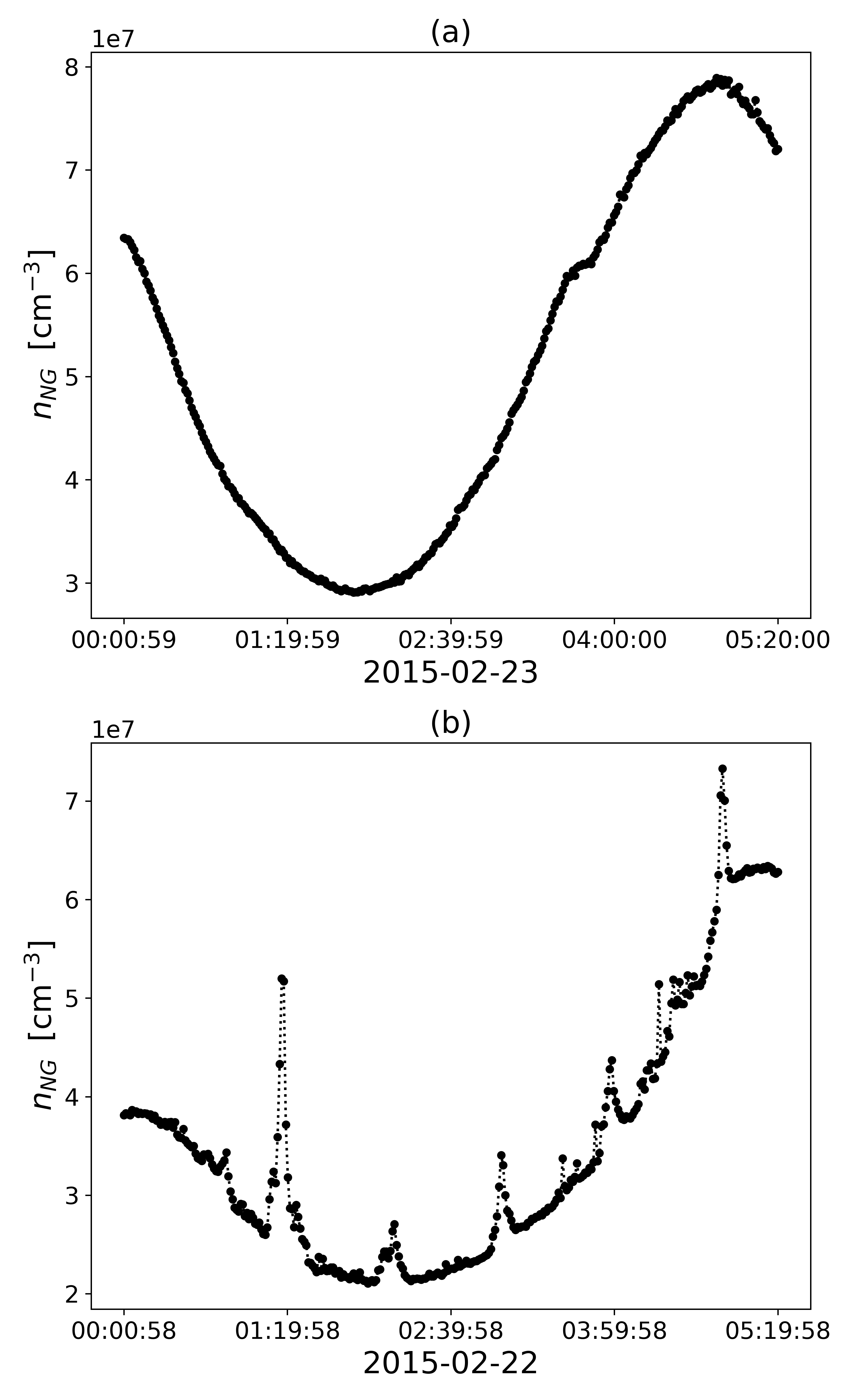}
\caption{Density measured by the NG. Dots represent actual measurements, whereas the dotted line connects the measurements to better visualise the trend. \textit{Panel a}: Measurements on 2015-02-23 between 00:00:59 and 05:20:00 without sublimating icy particle signatures. \textit{Panel b}: Measurements on 2015-02-22 between 00:00:58 and 05:19:58 with sublimating icy particle signatures.}
\label{fig:motivation}
\end{figure}

COPS is composed of two gauges: the aforementioned RG and the nude gauge (NG). The RG measures the ram pressure (gas flux), whereas the NG is dedicated to the measurement of the total neutral gas density.
This work demonstrates that the NG can also be used to study the volatile content of icy particles. In Fig.~\ref{fig:motivation}, we show the density, normalised to molecular nitrogen $\mathrm{N}_2$, measured by the NG over two consecutive days in February 2015. Unlike the relatively smooth and regular pattern in Fig.~\ref{fig:motivation}a that is typical for the diurnal evolution of the outgassing from the comet, Fig.~\ref{fig:motivation}b exhibits a series of sudden and short-lived density increases. As detailed in Part I, these features are generated by the sublimation of the volatile content of icy particles entering the gauge which was shown for the 5 September 2016 dust event \citep{Altwegg_et_al_2017}.

The nominal NG signal is the nominal density background, produced by gaseous coma species, measured by the NG (with the density increases generated by the sublimating content of icy particles totally removed). The nominal NG signal is lower and smoother than that of the RG, allowing the detection of significantly more features over the course of the mission. However, due to the different design of the two gauges \citep{Balsiger_et_al_2007}, the features produced by the volatile content of the icy particles in the NG data have a shorter duration (usually from one to six minutes) than those present in the RG data. This is because the NG, in its open design, has no physical barriers preventing the volatiles from escaping (Sect.~\ref{subsec:cops_nude_gauge_principle}). Since it is not possible to observe the decrease in density back to nominal NG values, as in Part I, the only value that can be directly extracted from the measured data is the feature height. Nevertheless, it is possible to perform a statistical analysis by combining the feature height with other parameters, such as the date, the sensor orientation with respect to the comet, the cometocentric distance (i.e. the distance of the spacecraft from 67P), and the heliocentric distance (i.e. the distance of the comet from the Sun). 

The methodology employed in this study is expounded on in Sect.~\ref{sec:methods} and the results of the analysis are presented in Sect.~\ref{sec:results}. A discussion of the caveats and complications is provided in Sect.~\ref{sec:discussion} and Sect.~\ref{sec:conclusion} summarises the findings and provides an outlook for future works.

%%%%%%%%%%%%%%%%%%
%%%%%%%%%%%%%%%%%%
%%%Methods
%%%%%%%%%%%%%%%%%%
%%%%%%%%%%%%%%%%%%

\section{Methods}\label{sec:methods}

\subsection{COPS nude gauge measuring principle}
\label{subsec:cops_nude_gauge_principle}
\begin{figure}
\centering
\includegraphics[width=0.3\textwidth]{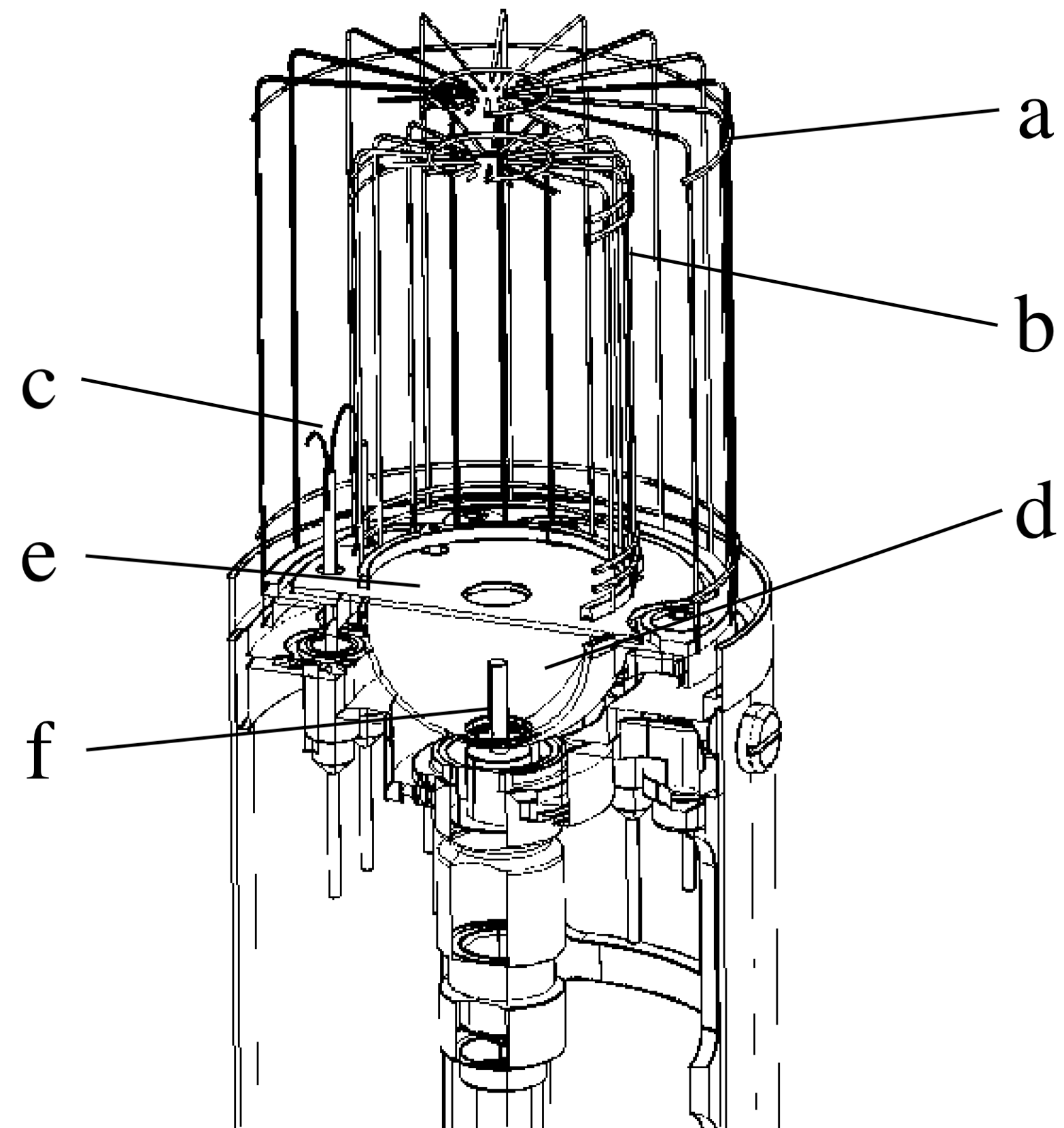}
\caption{COPS nude gauge. Main components: (a) outer grid kept at -12 V; (b) cylindrical (11 mm radius and 34 mm height) inner grid maintained at +180 V; (c) 17 mm long hot filament and its backup at +28 V potential; (d) hemispherical reflector at +110 V; (e) base plate at the potential of 0 V; and (f) ion collector.}
\label{fig:nude_gauge}
\end{figure}

The NG is the component of COPS devoted to the measurement of the total neutral density. This device is schematically represented in Fig.~\ref{fig:nude_gauge}. The gauge is constructed in such a way that gas and icy particles can enter from almost all directions. By default, the NG points in the -y direction of the Rosetta spacecraft, which points towards the solar panels; this means that cometary gas and dust are usually encountered from the side.

The instrument was built on the basis of the extractor-type ionisation gauge principle \citep[][]{Redhead_1966}. The outer (a) and inner (b) grids, respectively, prevent low-energy electrons and cometary ions from entering the volume enclosed within the inner grid. Consequently, this part of the instrument, known as the ionisation region, is only accessible to neutral particles. The neutral particles inside this region are ionised by free $\sim$150 eV electrons emitted by the filament (c) and accelerated towards the inner grid. During NG operations, only one of the two filaments was used, the other held in reserve. Atoms and molecules thus ionised were accelerated towards the base plate (e) and focused by the reflector (d) onto the ion collector (f). Since the ratio of the ion current measured by the collector to the filament's electron emission current is directly proportional to the density of neutral particles within the ionisation volume, it is possible to calculate the ambient particle density from the measured ion current. The relevant proportionality constants were determined from calibration experiments conducted in the laboratory with a 
``twin'' instrument identical to the one in space \citep[][]{Graf_et_al_2004}.

\subsection{Feature selection}
\label{subsec:features_selection}
The NG carried out almost one million measurements, one every minute, over the course of the mission. Here, we show that the features in the measured NG density should be carefully selected as some of them are not generated by the sublimation of the volatile content of icy particles.

The features were identified with a {\fontfamily{qcr}\selectfont Python} script, wherein $3\times10^5\,\mathrm{cm}^{-3}$ was defined as the minimum density above the nominal NG signal for a feature to be considered as such. This value was carefully selected so as to include as many features as possible, while simultaneously excluding the small fluctuations that represent the instrument noise of COPS, such as those present between 04:00:00 and 05:20:00 in Fig.~\ref{fig:motivation}a. The definition of $3\times10^5\,\mathrm{cm}^{-3}$ as the lower limit also enables the identification of features that at first glance do not have a clear peak shape, as the difference between their amplitudes and the nominal NG coma signal is not as prominent. As stated in Part I, these features are generated by the sublimation of volatiles in icy particles. 

\begin{figure*}
\centering
\includegraphics[width=\hsize]{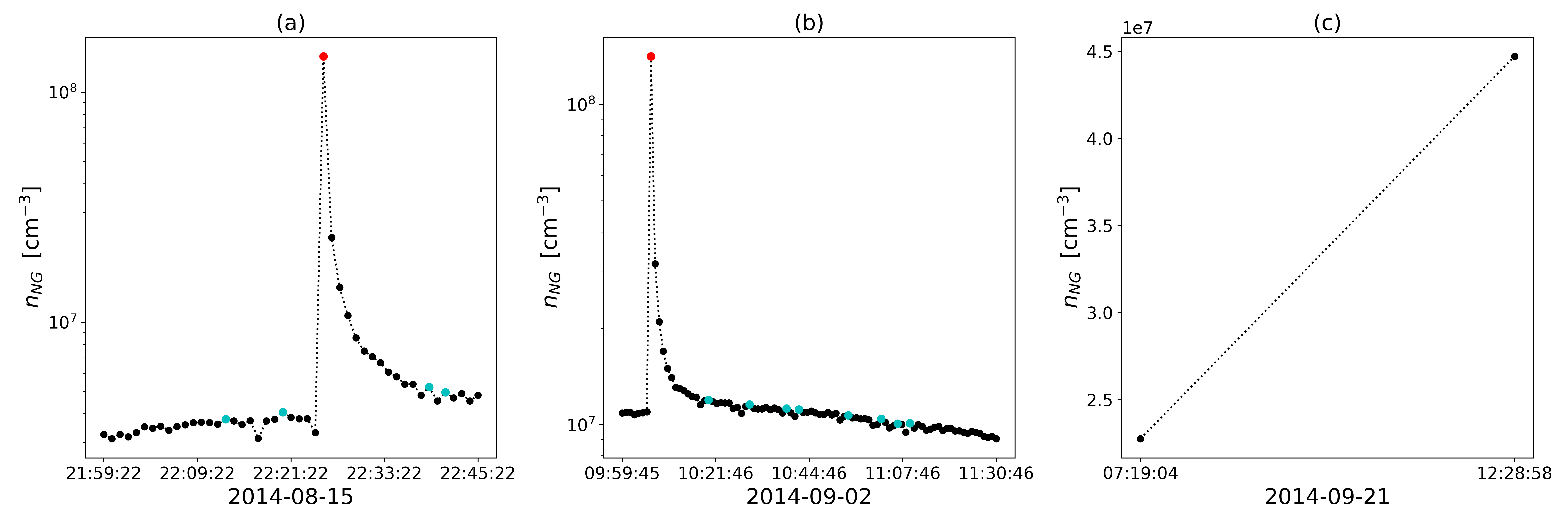}
\caption{Examples of NG density measurements during the three thruster operation periods. Dots represent actual measurements, whereas the dotted line connecting individual measurements acts as a reference to better visualise the changes in the density. The cyan dots indicate features generated by the sublimation of the volatile content of icy particles, whereas the red dots denote the features generated by thruster firings. \textit{Panel a}: Wheel offloading and navigation manoeuvre (MWNV) executed between 21:59:22 and 22:45:22 on 2014-08-15. The plot contains four features generated by the volatile content of icy particles and one (big) feature caused by the thrusters. \textit{Panel b}: Wheel offloading manoeuvre (MWOL) executed between 09:59:45 and 11:30:46 on 2014-09-02. The graph contains a thruster feature and eight features generated by the volatile content of icy coma particles. \textit{Panel c}: Orbital correction manoeuvre (MOCM) performed from 07:19:04 to 12:28:58 on 2014-09-21. In this time interval, no measurements were taken by the NG for about five hours.}
\label{fig:operation_modes}
\end{figure*}

Features that were measured during spacecraft manoeuvres were also excluded, as such manoeuvres often produced sudden increases in density associated with thruster operations that can be misinterpreted as coming from the sublimation of the volatile content of icy particles \citep{Schlappi_et_al_2010}. To distinguish the features produced by the volatile content of icy particles from those produced by thruster operations, the latter were characterised by analysing the $\sim$1800 periods dedicated to manoeuvres throughout the mission. In Fig.~\ref{fig:operation_modes}, examples of the density profiles produced by each of the three possible thruster operation modes are shown. Such manoeuvres lead to increases in density not caused by the volatile content of icy particles (red dots). However, these periods also contain features generated by sublimating volatiles coming from 67P that are relevant to this study (cyan dots). 

Certain features can also be caused by the sublimation of ices (mainly $\mathrm{H_2O}$) that have adsorbed onto the colder surfaces of the spacecraft itself \citep[][]{Schlappi_et_al_2010}. These are characterised by features coinciding with changes in spacecraft attitude and, consequently, illumination conditions, so they have also been discarded. 

Finally, any features that were measured during periods when high-energy electrons ($\gtrsim200$ eV) were detected by the Ion and Electron Sensor \citep[IES,][]{Burch_et_al_2007} of the Rosetta Plasma Consortium \citep[RPC,][]{Carr_et_al_2007} must be removed. This is because these features are not generated by the volatile content of icy particles, but are associated with the interaction of the solar wind with the plasma environment around Rosetta and COPS itself.  

Upon applying these criteria, the number of features (and equivalently icy particles) identified amounted to $\sim$67000. It is assumed that each feature corresponds to one icy particle, and thus that the NG was not measuring several icy particles at the same time. Since there is no way to determine how many icy particles are actually being detected in a single measurement -- only that there is at least one if not more -- the total number of events serves as a lower limit. This excludes possible showers of particles impacting the gauge nearly simultaneously, as given in the analysis in \citet{Fulle_2015}.

\subsection{Assumptions applied in the statistical analysis}
\label{subsec:stat_framework}
The comet is considered a single entity because there is no way to determine with certainty which part of the surface the icy particles that are detected by the NG have originated from. There are four reasons for this, the first being the unknown velocity and direction of the impacting icy particle, since NG data alone does not contain sufficient information to trace the particle to its source region as was done, for instance, by \citet{Longobardo_et_al_2020} using the GIADA measurements. The situation is further complicated by both the dust fallback phenomenon and the corresponding mass transfer from one region to another \citep[][]{Keller_et_al_2017}, as well as the large field of view that characterises the NG (Fig.~\ref{fig:nude_gauge}).
Finally, even if the volatile mass of an icy particle would be known (Sect.~\ref{subsec:cops_estimation_size_proc}), no inference about its amount of refractory material can be made and, therefore, the total mass and volume of the particle still remain unknown (see Sect.~\ref{subsec:dust_size}).

The temporal analysis utilised monthly intervals over which each quantity was averaged, so as to minimise the effects of short term variations arising from the rotation of 67P and the motion of the spacecraft around the comet \citep[][]{Hansen_et_al_2016}.

\subsection{Estimation of the volume of the volatile content of icy particles}
\label{subsec:cops_estimation_size_proc}
To estimate the amount of volatiles in the icy particles detected by the NG, a primarily geometric approach was employed. By taking advantage of the measurement principle presented in Sect.~\ref{subsec:cops_nude_gauge_principle}, and based on a number of assumptions, the quantity of volatiles within the icy particles detected by the NG can be estimated. The results are presented as the equivalent diameter of a spherical particle of density 1 g cm$^{-3}$, close to a pure water ice particle without porosity.
Similarly to the observations in Part I for the RG, it is possible that the NG-detected icy particles are comprised purely of volatiles. However, no conclusion can be drawn regarding the refractory content of the icy particles as this component is invisible to both the NG and the RG. This issue will be explored in more detail in Sect.~\ref{sec:discussion}.

It is assumed that when an icy particle enters the ionisation volume of the NG, its volatile content will be distributed evenly throughout the ionisation volume. Thus, it is possible to obtain the number of volatile molecules $N_\mathrm{vol}$ by multiplying the volume of the ionisation region $V_\mathrm{ion}$ 
%$V_{ion}=(11\mathrm{mm})^2\cdot\pi\cdot34\mathrm{mm}$ 
by the height of the feature $n_\mathrm{dust}$, which is the total density minus the nominal coma density: $N_\mathrm{vol}=V_\mathrm{ion}\, n_\mathrm{dust}$. By considering only the height, which is the highest data point in a feature following the subtraction of the nominal NG signal, it is assumed that all the volatile content of an icy particle is sublimated during each single measurement by the NG. The amounts of volatiles in particles that require several minutes to undergo sublimation are therefore underestimated. This results in a lower limit of the total amount of volatiles for those icy particles that generated a feature spanning multiple measurements. It was shown in Part I that icy particles may be classified into multiple groups that are characterised by different volatile compositions. However, the COPS data cannot give any indication regarding the molecular composition of the different types of icy particles. Therefore, given the fact that water is the most abundant cometary volatile \citep[][]{Le_roy_et_al_2015,Snodgrass_et_al_2016}, a water-like species is assumed as the sole constituent of the volatiles for our size estimation. For simplicity, since no information about the type of water ice that constitutes the icy particles is available, the density of liquid water was used (1 g cm$^{-3}$), but the results can easily be scaled to any other density. The volatile volume $V_\mathrm{vol}$ can thus be estimated by multiplying the number of volatile molecules by the volume of one $\mathrm{H_2O}$ molecule, as $V_\mathrm{vol}=N_\mathrm{vol}\, V_\mathrm{H_2O}$. The diameter $d$ of a spherical particle having a volume equivalent to $V_\mathrm{vol}$ can be expressed as:
\begin{equation}
\label{eqn:geom_diameter_volatiles}
d = 2\sqrt[3]{\frac{3}{4\pi}V_\mathrm{H_2O}\, V_\mathrm{ion} \, n_\mathrm{dust}} = 0.9\,\sqrt[3]{n_\mathrm{dust}[\mathrm{cm^{-3}}]} \;\mathrm{nm}.
\end{equation}
The values obtained from this calculation are lower limits, as it is assumed that all volatiles from the icy particles were within the ionisation region of the NG and the feature amplitude instead of the area was used. 
It is not possible to determine whether the NG was indeed measuring all or only a fraction of the volatile content of icy particles. This will be addressed in greater detail in Sect.~\ref{sec:discussion}.

%%%%%%%%%%%%%%%%%%
%%%%%%%%%%%%%%%%%%
%%%Results
%%%%%%%%%%%%%%%%%%
%%%%%%%%%%%%%%%%%%

\section{Results}\label{sec:results}

The results of the analyses performed are presented as follows: 1) a parameter variation analysis of the features (Sect.~\ref{subsec:param_varia_analysis}); 2) an investigation of the temporal distribution of the detected icy particles (Sect.~\ref{subsec:counts_and_event_rate}); 3) a comparison between the NG detection rate (i.e. the rate of icy particles detected by the NG) and quantities associated with cometary activity such as the water production rate and the ground-based dust brightness measurements (Sect.~\ref{subsec:dust_comparison}); and 4) a calculation of the size distribution of the volatile content of the icy particles (Sect.~\ref{subsec:size_distribution}).

\subsection{Parameter variation analysis}
\label{subsec:param_varia_analysis}
\begin{figure*}
\centering
\includegraphics[width=0.75\textwidth]{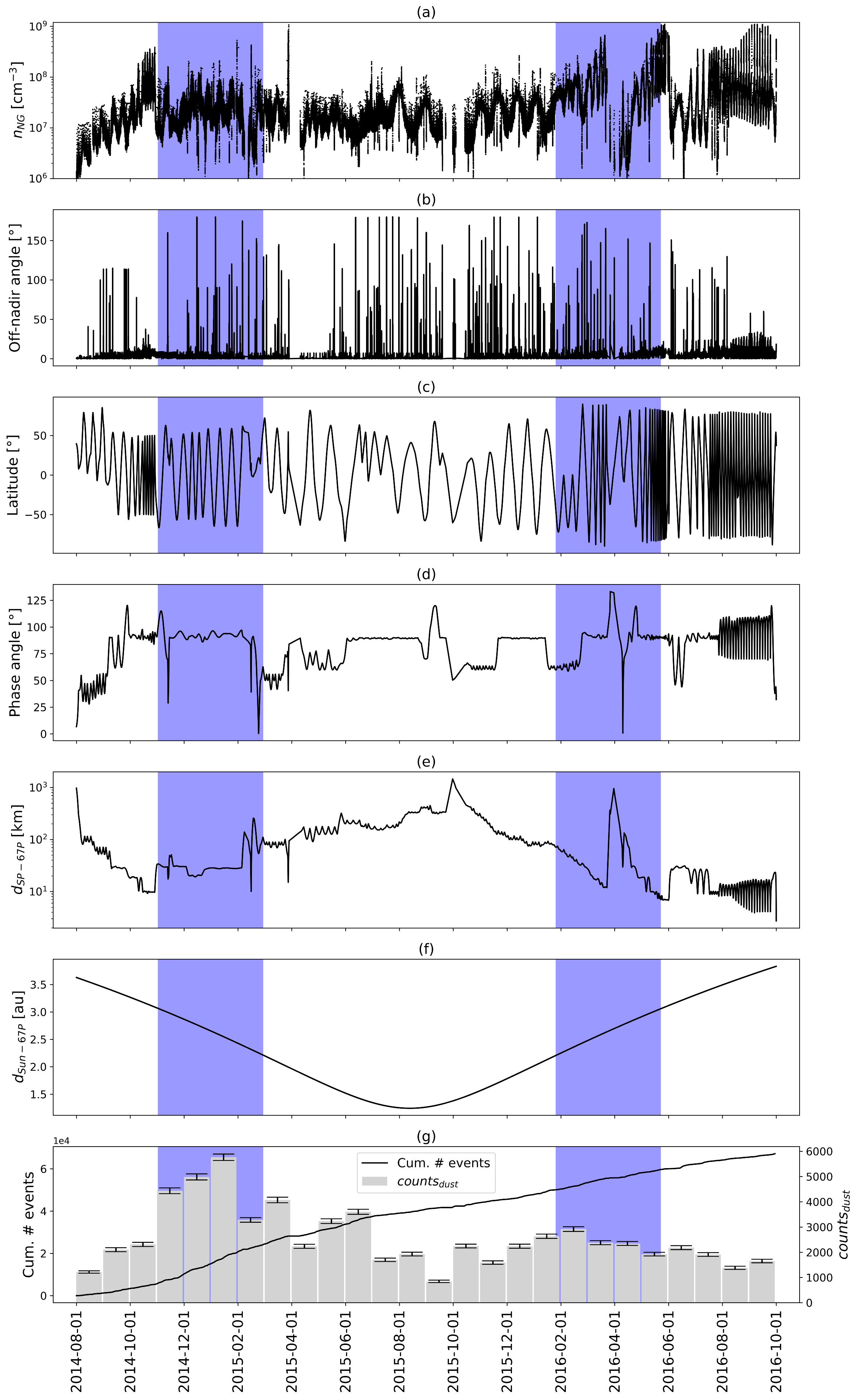}
\caption{Six parameters investigated in this study, together with the number of detections, plotted as a function of time. From top to bottom: (a) density (nominal NG signal); (b) off-nadir angle; (c) (sub-spacecraft) latitude; (d) phase angle; (e) cometocentric distance; (f) heliocentric distance; (g) cumulative number of NG detection events and number of icy particles detected by the NG per month (the error bars represent the statistical uncertainty). The blue bands indicate the periods that are investigated in Sect.~\ref{subsubsec:selected_interval_disentangl}.}
\label{fig:param_overview}
\end{figure*}

Icy particles detected by the NG can be classified according to the parameters that were used in their measurements. Six different parameters are considered in this work and the dependence of the results on each of them should be assessed separately.
Figure~\ref{fig:param_overview} shows the six parameters of interest in this study -- namely, the density (nominal NG signal), off-nadir angle, (sub-spacecraft) latitude, phase angle, cometocentric distance, and heliocentric distance -- as well as the number of detections. As can be seen from Fig.~\ref{fig:param_overview}g, detections of the icy particles occurred throughout the entire course of the mission. Therefore, Fig.~\ref{fig:param_overview}a-f are also a function of the spacecraft's orientation during the detections.

The first parameter investigated is the nominal NG signal, which is the density without the features generated by the sublimation of the volatile content of an icy particle. The density measured by the NG spans a range from $2.7\times10^5\,\mathrm{cm}^{-3}$ to $1.1\times10^9\,\mathrm{cm}^{-3}$ (see Fig.~\ref{fig:param_overview}a). The majority of the detections of icy particles ($\sim$90\%) were observed while the surrounding coma density was at the lower end of this range ($<5.0\times10^7\,\mathrm{cm}^{-3}$). Higher nominal NG signals also tended to coincide with fewer detections. This is because the heights of features generated by icy particles are usually between $10^5\,\mathrm{cm}^{-3}$ and $10^8\,\mathrm{cm}^{-3}$ and are harder to observe when the nominal NG signal is high at the time they blend into it. Rosetta was surrounded by a gaseous background of $\gtrsim2.5\times10^5\,\mathrm{cm}^{-3}$ \citep{Schlappi_et_al_2010}.

The off-nadir angle corresponds to the deviation of the Rosetta z-axis from the direction of the centre of mass of 67P. Most of the time, Rosetta was pointing at or close to the comet's nucleus. Consequently, $\sim$90\% of the detections have an off-nadir angle of less than 9$^\circ$. The remaining 10\% of icy particle detections have off-nadir angles between $10^\circ$ and $180^\circ$, with a slight tendency towards there being fewer events at larger angles (compare Fig.~\ref{fig:param_overview}b and Fig.~\ref{fig:param_overview}g). This is to be expected, given that very few measurements have been taken at the larger off-nadir angles. However, the NG detection rate, calculated by dividing the number of detections by the time spent at their corresponding angle, is almost constant for every range of off-nadir angles considered, indicating that this parameter has no significant influence on the NG detection rate. This is unsurprising, considering the large field-of-view of the NG.

The sub-spacecraft latitude of Rosetta is the third parameter considered (see Fig.~\ref{fig:param_overview}c). The NG detection rate (no.\ of detections divided by latitude) does not appear to depend on this parameter either, as it remains more or less constant regardless of the latitude. However, the absolute number of icy particle detections tends to be higher at latitudes between $-60^\circ$ and $60^\circ$. This can be explained by considering how little time Rosetta spent at latitudes that were further away from the comet equator.

The dependence of the results on the first three parameters -- off-nadir angle, nominal NG signal, and sub-spacecraft latitude -- is unambiguous. This is not the case for the phase angle, which is defined as the angle between 67P, the Rosetta spacecraft and the Sun. There is no single range of phase angles into which most of the icy particle detections fall, as can be seen by comparing Fig.~\ref{fig:param_overview}d and Fig.~\ref{fig:param_overview}g. 

For the cometocentric distance and the heliocentric distance, the treatment is also more complicated due to a complex interplay between these two parameters. To avoid issues associated with sunlight reflection off dust interfering with Rosetta's star trackers, the cometocentric distance was generally larger when 67P was close to its perihelion and peak activity, and decreased when 67P was at greater heliocentric distances (see Fig.~\ref{fig:param_overview}e and Fig.~\ref{fig:param_overview}f). This hinders any attempt to evaluate them separately. The combined influence of these two parameters is thus investigated by considering a limited time interval from the mission.

\subsubsection{Analysis of data from 2014-11-01 to 2015-02-28}
\label{subsubsec:selected_interval_disentangl}
To investigate the possible dependence of the results on the heliocentric distance and the cometocentric distance, this subsection focuses on the period between 2014-11-01 and 2015-02-28 during 67P's inbound trajectory. This interval was chosen for its potentially better statistics, since there were more detected icy particles during this time than in any other period, as can be seen from Fig.~\ref{fig:param_overview}g. A comparison with the outbound period at the same heliocentric distance will follow.

The influence of the off-nadir angle, nominal NG signal, and latitude on icy particle detection in this interval is first compared with the average for the entire mission from Sect.~\ref{subsec:param_varia_analysis}. The results are in agreement: 96\% of the icy particles were detected at a nominal NG signal lower than $5.0\times10^7\,\mathrm{cm}^{-3}$, 94\% of them were detected at an off-nadir angle lower than $9^\circ$, and the latitude has no observable influence on the distribution of the detections. 

To simplify the analysis by restricting the dependence of the results on the aforementioned parameters, only the icy particles detected at a nominal NG coma signal lower than $5.0\times10^7\,\mathrm{cm}^{-3}$ and at an off-nadir angle lower than $9^\circ$ were selected for the next step. These icy particles were investigated for their dependence on the phase angle. The range containing $\sim$80\% of the events was examined and 81\% of the detections are found to have a phase angle in the range between $86.3^\circ$ and $97.8^\circ$. This result is expected considering that observations were made at a phase angle consistently close to $90^\circ$ for most of this interval, with a few prolonged exceptions near the beginning and end of the investigated time period, as can be seen from Fig.~\ref{fig:param_overview}d. 

\begin{figure}
\centering
\includegraphics[width=\hsize]{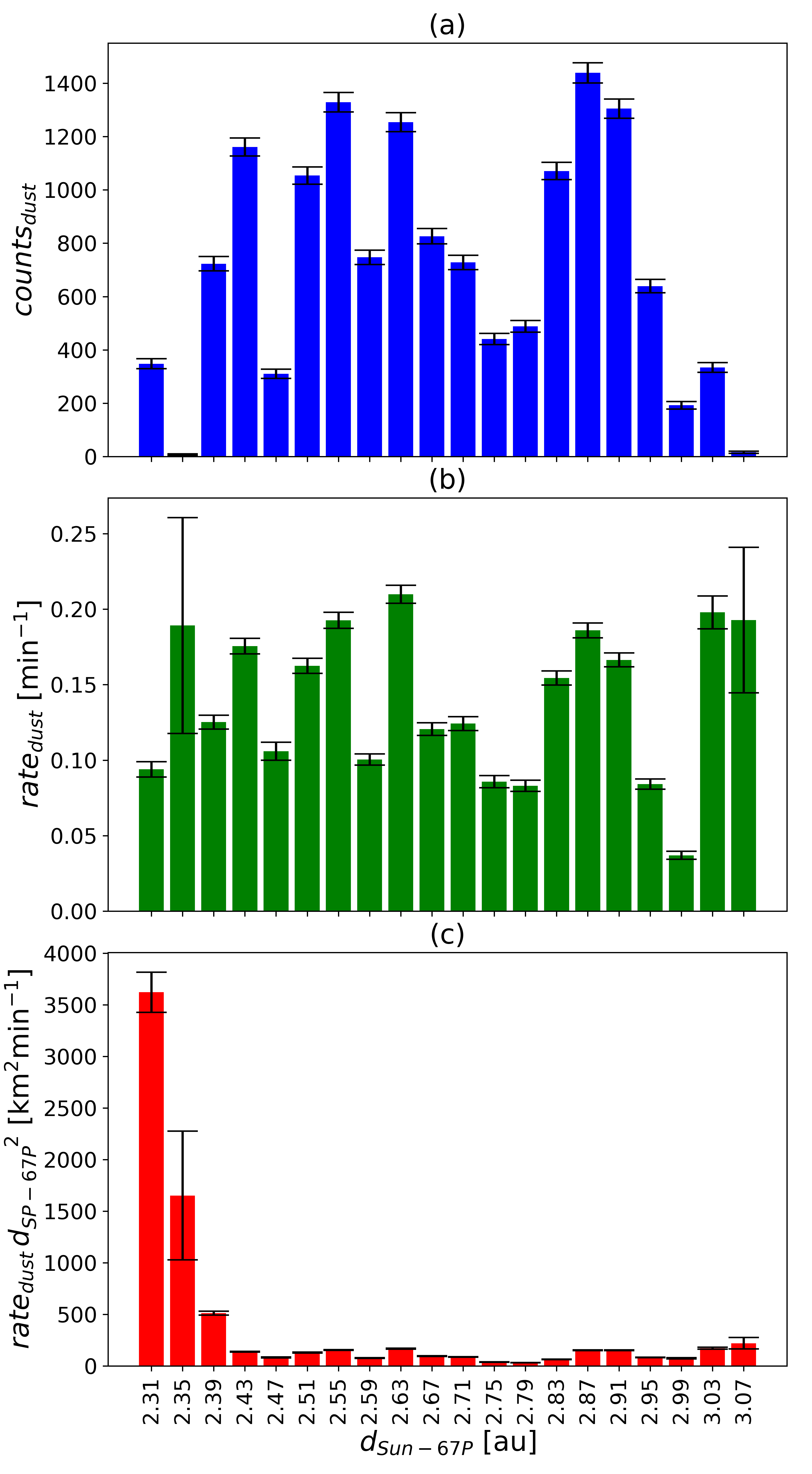}
\caption{Icy particles analysed were measured between 2014-11-01 and 2015-02-28, at a nominal NG signal less than $5.0\times10^7\,\mathrm{cm}^{-3}$, an off-nadir angle lower than $9^\circ$, and a phase angle in the range between $86.3^\circ$ and $97.8^\circ$. On the x-axis, the central values of bins with a width of 0.04 au are shown. \textit{Panel a}: Number of icy particle detections as a function of the heliocentric distance. \textit{Panel b}: NG detection rate as a function of heliocentric distance. \textit{Panel c}: NG detection rate, multiplied by the square of the average cometocentric distance, as a function of the heliocentric distance.}
\label{fig:disentanglement}
\end{figure}

Detections with a phase angle between $86.3^\circ$ and $97.8^\circ$ were then analysed to determine the nature of their dependence on the cometocentric distance and the heliocentric distance. As previously noted, it is not possible to analyse these two parameters separately and they must therefore be considered together. In Fig.~\ref{fig:disentanglement}, three quantities have been plotted as functions of the heliocentric distance. These are: (a) the number of icy particle detections; (b) the rate at which they occur; and (c) the detection rate multiplied by the square of the average cometocentric distance. This implicitly assumes that the icy particles reaching Rosetta (at distances >10 km) will ultimately escape and not fall back onto the surface of the nucleus.
No clear trends are discernible in Fig.~\ref{fig:disentanglement}a and Fig.~\ref{fig:disentanglement}b. However, Fig.~\ref{fig:disentanglement}c shows that if the data are also weighted based on the square of the average cometocentric distance, a clear trend emerges: the closer the comet is to the Sun, the higher the NG detection rate. The same pattern appears when cometary activity is considered, as this quantity also increases as 67P approaches the Sun \citep[][]{Hansen_et_al_2016,Biver_2019}.
This suggests a possible connection between cometary activity and the NG detection rate. 

The interval analysed corresponds roughly to a heliocentric distance of 3.07 to 2.21 au during the comet's inbound trajectory. An analysis of the interval corresponding to the same heliocentric distance during the outbound trajectory reveals the same pattern as the one shown in Fig.~\ref{fig:disentanglement}. The only difference is the lower number of icy particles detected by the NG between 2016-01-25 and 2016-05-23 (see Fig.~\ref{fig:param_overview}g). This and other aspects of the temporal distribution of the detections are addressed in the next subsection.

\subsection{Counts and NG detection rate}
\label{subsec:counts_and_event_rate}
From the histogram in Fig.~\ref{fig:param_overview}g, it can be seen that there were approximately 2000 detections per month, except for a few months in which only about 1000 icy particles were detected, such as August 2014 and September 2015, and the months between November 2014 and June 2015 (excluding April 2015) when more than 3000 icy particles were detected per month.

Based on the number of icy particles detected by the NG per month, there appear to be many more detections at the beginning of the mission. The frequent occurrence of features in the months between the end of 2014 and the beginning of 2015, relative to the period corresponding to the same heliocentric distance during the outbound orbit, could mean that there were more frozen volatiles (i.e. more ice-containing particles) during the inbound orbit than the outbound orbit at the same heliocentric distances. It is not possible for the higher number of detections at the beginning of the mission to be related to the cometocentric distance, given that Rosetta also spent extended periods at comparable distances or even closer to the nucleus at the end of the mission (see Fig.~\ref{fig:param_overview}e). We hypothesise that the higher number of detections during the comet's inbound trajectory as compared to its outbound one may be a reflection of the evolving properties of the nucleus. The surface of the nucleus was covered in a dust layer during the inbound part of the orbit, which was shed as the comet approached perihelion, potentially leading to a higher flux of icy particles than during the outbound part of the orbit when no extra layer of dust was present on the surface \citep{Guilbert_Lepoutre_et_al_2014,Schulz_et_al_2015}. The temporal distribution of icy particles detected by the NG suggests, therefore, that the cover of dust proposed by \citet{Schulz_et_al_2015} can be extended to the volatile content. This trend was also observed in Part I in data from the RG, but no conclusions were drawn. This was due to possible bias caused by the lower detection statistics and the reduction in sensitivity of the RG in the middle of the mission due to a change in the mode of operation. Based on a comparison of the measurements from the two COPS gauges, however, it can now be confirmed that the higher number of detections observed at the beginning of the mission by the RG is indeed present in the NG measurements as well.

\begin{figure*}
\centering
\includegraphics[width=\hsize]{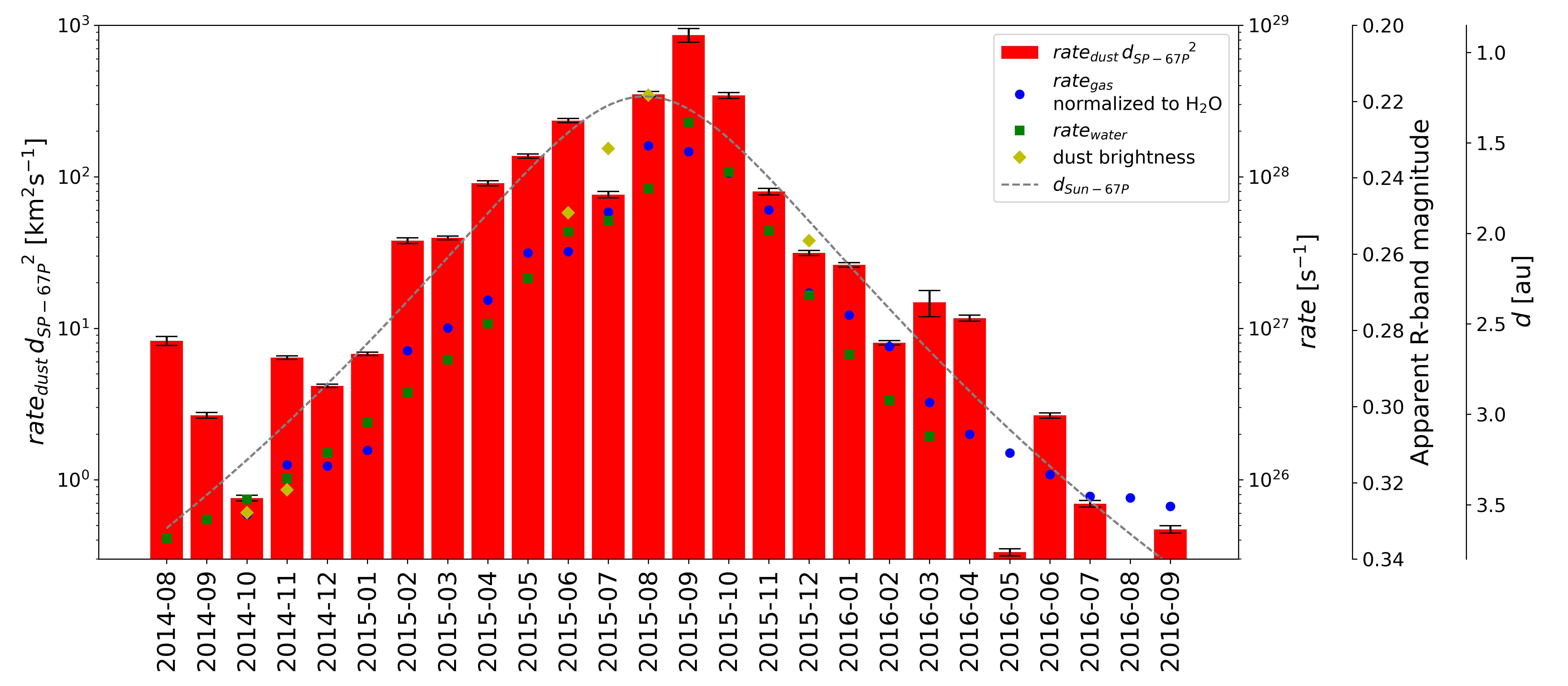}
\caption{Comparison of histogram of the NG detection rate multiplied by the square of the average cometocentric distance (red bars), with the monthly averages of the gas production rate $Q$ (normalised to $\mathrm{H_2O}$) calculated from the NG data (blue dots), the water production rate from \citet[Figure~9]{Hansen_et_al_2016} (green squares), the ground-based dust brightness observations of \citet{Snodgrass_et_al_2016} (yellow diamonds), and the heliocentric distance (grey dashed line).}
\label{fig:event_rate}
\end{figure*}

As noted in Sect.~\ref{subsec:param_varia_analysis}, the NG measurements were taken under changing conditions. To draw additional conclusions from the temporal distribution of icy particles detected by the NG over the course of the mission, the number of detections must be weighted based on both the measurement time (to correct for differences in exposure times each month) and cometocentric distance, as in Fig.~\ref{fig:disentanglement}c. In Fig.~\ref{fig:event_rate}, the number of detections per month from Fig.~\ref{fig:param_overview}g has been both divided by the measurement time and multiplied by the square of the average cometocentric distance. It can be seen from Fig.~\ref{fig:event_rate} that the months with higher NG detection rates are the ones around mid-2015. As the heliocentric distance (grey dashed line in Fig.~\ref{fig:event_rate}) shows, this was when 67P reached its perihelion and was thus more active \citep[][]{Hansen_et_al_2016}. This again suggests the existence of a positive correlation between the NG detection rate and cometary activity.

\subsection{Comparison of NG detection rate with gas production rate, water production rate, and ground-based dust brightness measurements}
\label{subsec:dust_comparison}
The results of the preceding subsections hint at a possible connection between the NG detection rate and cometary activity. This subsection investigates this potential connection by looking for possible correlations between the NG detection rate multiplied by the square of the average cometocentric distance, and (i) the monthly averages of the total gas production rate, calculated by applying the Haser model to the NG data, (ii) the monthly averages of the water production rate presented in \citet{Hansen_et_al_2016}, as well as (iii) the monthly averages of the ground-based dust brightness measurements presented in \citet{Snodgrass_et_al_2016}. 

\begin{figure}
\centering
\includegraphics[width=\hsize]{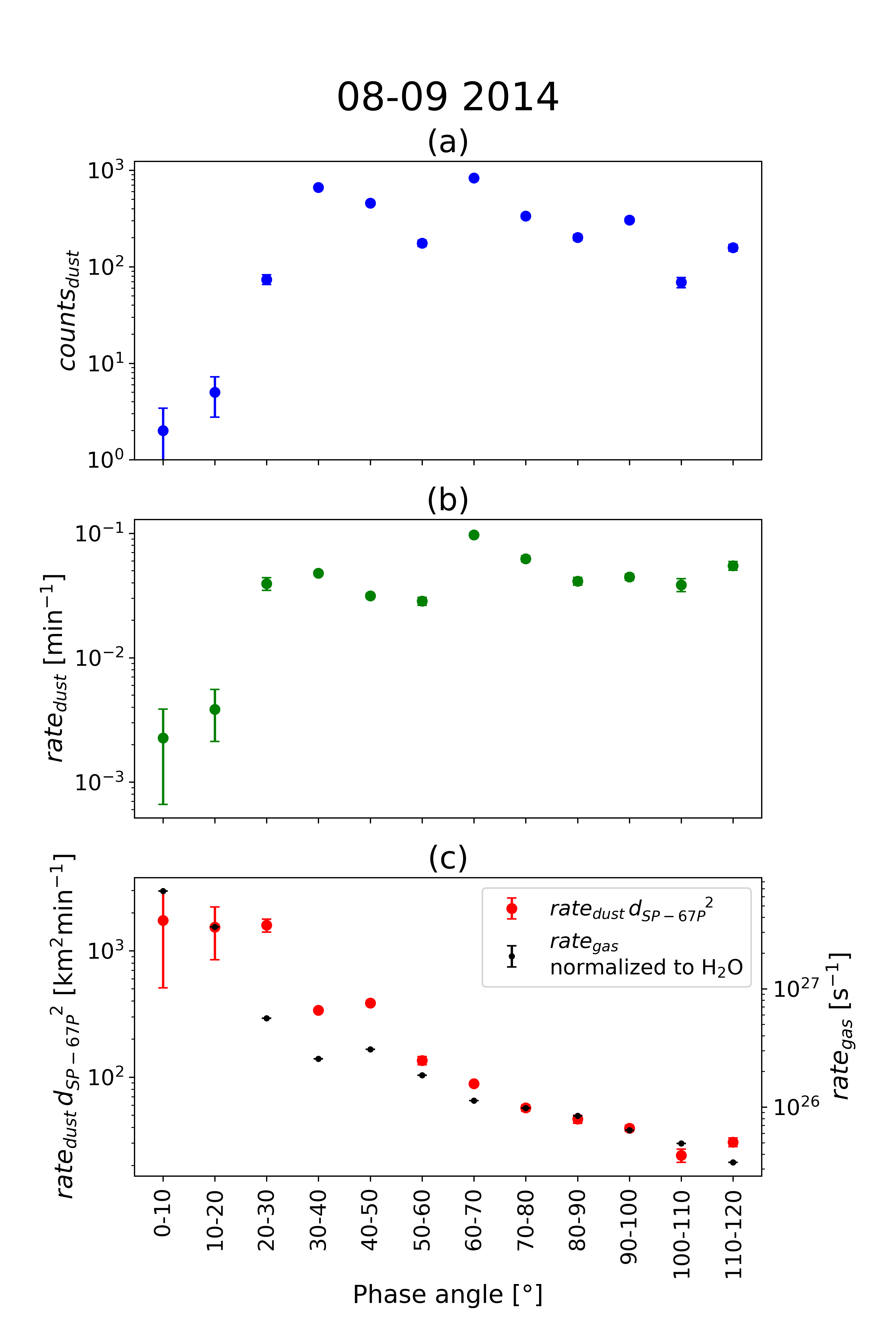}
\caption{Investigation of possible phase angle dependencies during August and September 2014. \textit{Panel a}: Number of icy particle detections. \textit{Panel b}: NG detection rate. \textit{Panel c}: NG detection rate multiplied by the square of the average cometocentric distance, compared to the gas production rate $Q$ normalised to $\mathrm{H_2O}$.}
\label{fig:phase_angle_dep}
\end{figure}

Figure~\ref{fig:event_rate} shows a superposition of the different datasets. It should be noted that the dust brightness is rather a measure for the amount of dust in the coma than the dust activity (i.e. dust production) of the comet. 
The total gas production rate $Q$ was estimated with the Haser model \citep[][]{Haser_1957}, which provides a reasonable order of magnitude approximation despite assuming radial outgassing and spherical symmetry \citep[][]{Rezac_et_al_2019}:
\begin{equation}
\label{eqn:haser}
Q\left(d_\mathrm{SP-67P}\right)=4\pi\, v\, n\left(d_\mathrm{SP-67P}\right) \, \left.d_\mathrm{SP-67P}\right.^2,
\end{equation}
where $d_\mathrm{SP-67P}$ is the cometocentric distance, $v$ is the gas velocity, and $n\left(d_\mathrm{SP-67P}\right)$ is the gas density measured by the NG. In this calculation, both the cometocentric distance and the measured NG density are averaged over one-month intervals. The production rate (green squares) modelled by \citet[]{Hansen_et_al_2016} corrects for the vantage point of spacecraft in terms of phase angle, sub-spacecraft latitude, and longitude. This is not taken into account in our Haser-type model (blue circles, cf. Eq.~\ref{eqn:haser}), which explains the difference between the two. For this reason -- and due to a combination of the low signal-to-noise ratio (S/N), small phase angles, and large cometocentric distances -- the production rate derived from the Haser model has been omitted for Rosetta's initial approach to the comet in August and September 2014.
In the first two months of the mission, detected icy particles at phase angles smaller than 20$^\circ$ are scarce (see Fig.~\ref{fig:phase_angle_dep}a) and the corresponding detection rates are low (see Fig.~\ref{fig:phase_angle_dep}b). This is due to a low S/N and a lack of measurements carried out at these phase angles. However, these detections were made at large cometocentric distances and, therefore, the resulting NG detection rate multiplied by the square of the average cometocentric distance is high (see Fig.~\ref{fig:phase_angle_dep}c). The few detections at phase angles smaller than 20$^\circ$ significantly affect both the NG detection rate multiplied by the square of the average cometocentric distance and the gas production rate (cf. Eq.~\ref{eqn:haser}). This explains our choice not to provide a Haser model-derived production rate for August and September 2014, and also why the NG detection rate multiplied by the square of the average cometocentric distance is relatively high for these two months (see Fig.~\ref{fig:event_rate}). In Fig.~\ref{fig:phase_angle_dep}c, the differences between the gas production rates at various phase angle intervals are greater than a factor of ten. This value is similar to the one derived by \citet[Fig. 2]{Hansen_et_al_2016} for comparable spacecraft orientations. Similarly to the gas, the dust also shows enhanced fluxes towards low phase angles \citep[cf.][]{Combi_2012}. However, the reader should be cautioned that our result is influenced by poor coverage at low phase angles.

The gas production rate exhibits a dichotomy with respect to 67P's perihelion: pre-perihelion, the NG detection rate was always higher than the gas production rate, whereas post-perihelion -- with the exception of a few months (March, April, and July 2016) -- the NG detection rate was either similar to or lower than the gas production rate. One possible explanation for this behaviour could again be the shedding of a mantle of dust during the approach towards perihelion.

\subsubsection{Comparison of NG detection rate and gas production rate}
\label{subsubsec:comparison_gas_rate}
Because the density measured is normalised to molecular nitrogen, the gas production rate calculated with Eq.~\ref{eqn:haser} is initially also normalised to N$_2$. Using a correction factor \citep[0.893,][]{granville_phillips_2007}, we then renormalised the gas production rate assuming 100\% H$_2$O. In this way, the gas production rate is directly related to the water production rate (see Sect.~\ref{subsubsec:comparison_water_rate}).
The assumed gas velocity is $v\simeq700\,\mathrm{m/s}$, in accordance with the average value recorded by MIRO, a microwave instrument on board Rosetta devoted to the measurement of $\mathrm{H_2O, CO, NH_3,}$ and $\mathrm{CH_3OH}$ \citep[][]{Gulkis_et_al_2015}. As previously stated, this approach is intrinsically approximated.

\begin{figure}
\centering
\includegraphics[width=0.4\textwidth]{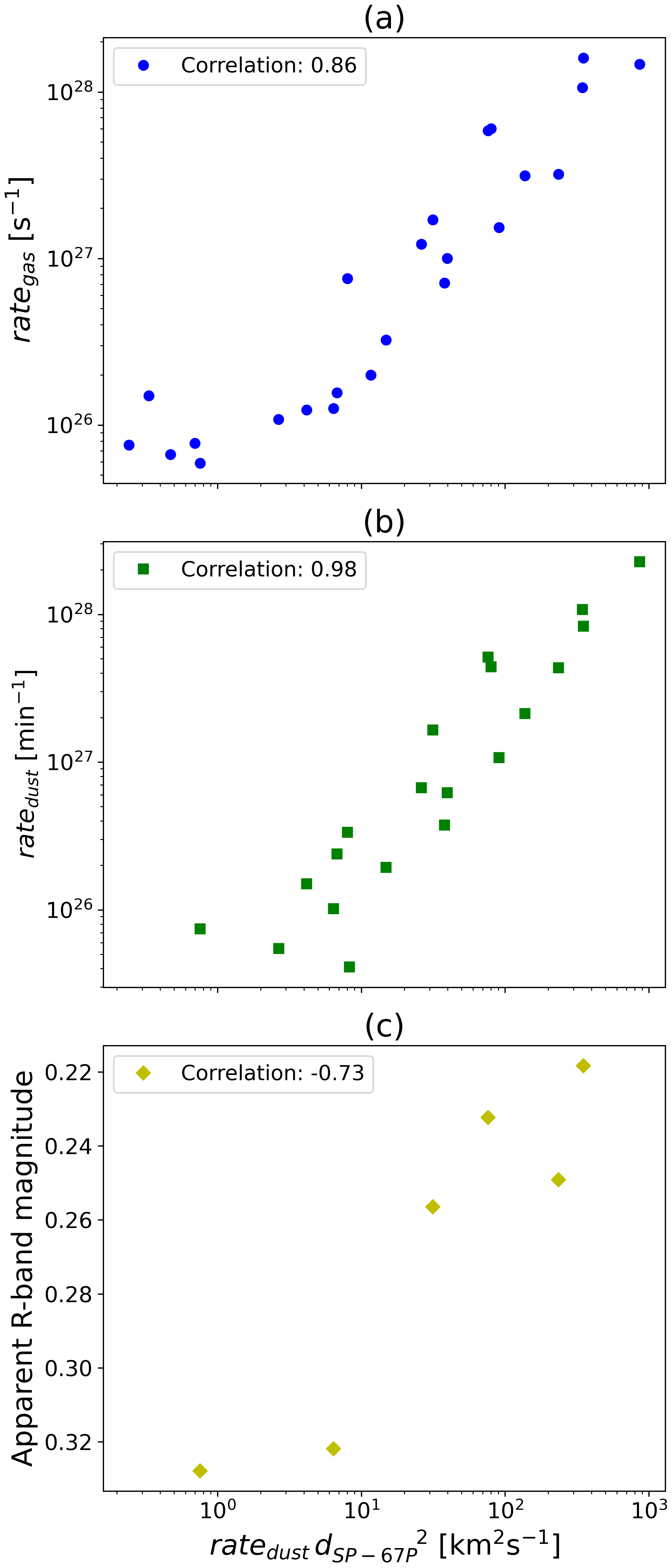}
\caption{Direct comparisons of the datasets from Fig.~\ref{fig:event_rate}. \textit{Panel a}: Monthly averages of the gas production rate $Q$ (derived from NG data) plotted against the monthly NG detection rate multiplied by the square of the average cometocentric distance. The Pearson correlation coefficient is 0.86. \textit{Panel b}: Monthly averages of the water production rate \citep{Hansen_et_al_2016} versus the monthly NG detection rate multiplied by the square of the average cometocentric distance. The Pearson correlation coefficient between the two quantities is 0.98. \textit{Panel c}: Monthly average of the ground-based measurements of the brightness of 67P \citep{Snodgrass_et_al_2016} versus the monthly NG detection rate multiplied by the square of the average cometocentric distance. The Pearson correlation coefficient is -0.73.}
\label{fig:comparisons}
\end{figure}

Based on the data in Fig.~\ref{fig:event_rate}, the monthly NG detection rate multiplied by the square of the average cometocentric distance was compared to the monthly averages of the gas production rate, $rate_\mathrm{gas}$ (Fig.~\ref{fig:comparisons}a). A particularly high Pearson correlation between them was found (0.86), while the p-value is very low ($6.3\times10^{-8}$). This confirms that the NG detection rate and the cometary activity are indeed connected. 

\subsubsection{Comparison of NG detection rate and water production rate}
\label{subsubsec:comparison_water_rate}
In a subsequent step, the monthly NG detection rate multiplied by the square of the average cometocentric distance was compared with the monthly averages of the empirically-corrected water production rate presented in \citet[]{Hansen_et_al_2016}. The result obtained (see Fig.~\ref{fig:comparisons}b) is very similar to that in Sect.~\ref{subsubsec:comparison_gas_rate}, but with an even higher Pearson correlation of 0.98 and a lower p-value of $2.5\times10^{-13}$, which may be related to the fact that the dataset being compared was based on a series of direct measurements of the comet, whereas the data in Fig.~\ref{fig:comparisons}a was based on a simplified theoretical model.

\subsubsection{Comparison of NG detection rate and ground-based dust brightness measurements}
\label{subsubsec:comparison_ground_based_measur}
The monthly averages of the dust brightness measurements were calculated and plotted against the monthly NG detection rate multiplied by the square of the average cometocentric distance (Fig.~\ref{fig:comparisons}c). Although Fig.~\ref{fig:comparisons}c contains only six data points, they exhibit an anti-correlation. The negative Pearson correlation (-0.73), with a p-value of $9.9\times10^{-2}$, comes about because lower magnitudes correspond to brighter objects. If 67P has more dust, it will elicit a higher NG detection rate, in addition to reflecting more light and thus having a lower magnitude in brightness. It should, however, be noted that the range of particle sizes measured by the NG may differ from those measured by remote observations, especially considering that the NG detects only volatiles, whereas \citet{Snodgrass_et_al_2016} observed both volatiles and refractories.

\subsection{Size distribution}
\label{subsec:size_distribution}
\begin{figure}
\centering
\includegraphics[width=\hsize]{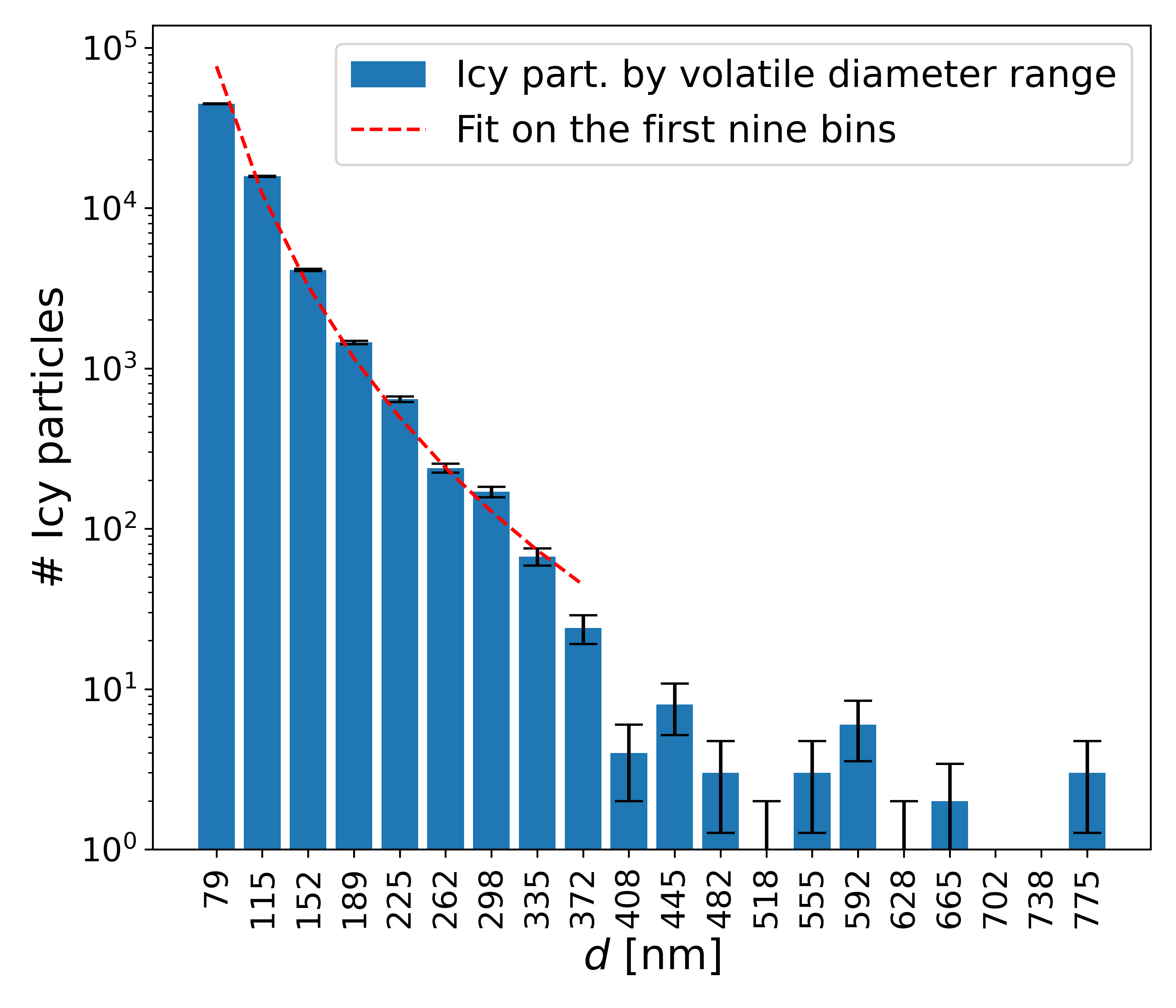}
\caption{Size distribution of the volatile content of the icy particles. The horizontal axis shows the diameters of spherical particles (density of 1 g cm$^{-3})$ having the same volume as the volatiles in the icy particles. The displayed x-axis values are the central values of bins with a width of $\sim$36.5$\,\mathrm{nm}$ (the width was chosen so as to have exactly 20 bins).
The values were calculated with Eq.~\ref{eqn:geom_diameter_volatiles} and the error bars represent the statistical error. The red dashed line is the best fit of the first nine bins according to Eq.~\ref{eqn:fit_size_dist}.}
\label{fig:size_distr}
\end{figure}

Figure~\ref{fig:size_distr} shows a histogram of the size distributions of the volatile content of icy particles calculated from the peak densities of the $\sim$67000 features described in Sect.~\ref{subsec:features_selection} (cf. Eq.~\ref{eqn:geom_diameter_volatiles}). The error bars indicate the statistical uncertainties. 

By assuming a density of 1 g cm$^{-3}$, the volatile part of the detected icy particles can be represented as equivalent spheres having a diameter in the range between $\sim$60 $\mathrm{nm}$ and $\sim$793 $\mathrm{nm}$. The refractory part of the icy particles is neglected (see Sect.~\ref{subsec:cops_estimation_size_proc}).

To assess the reliability of the results acquired with Eq.~\ref{eqn:geom_diameter_volatiles}, the diameters of the volatile content of the icy particles detected by the NG can be compared with those obtained with the RG in Part I. The RG provides, in addition to its measurement of the (ram) density, a much better indication of the volatile volume. This is because the RG had a closed design and, hence, particles sublimating inside it produced a sharp increase in density, followed by a slower decrease, a characteristic of the low-pressure effusion of gas. Sublimation is very efficient as COPS is much warmer than the surrounding coma (typically 0 degrees Celsius as measured for the RG, cf. Part I). The NG, in contrast, has no physical barriers, and volatiles can easily escape the ionisation region during a measurement. On the other hand, the different construction of the two gauges also means that the detection rate of the NG is higher than that of the RG.
For the comparison between icy particles detected by the NG and RG, it is assumed that these are similar with respect to composition and, therefore, that their amount of volatiles should also be alike. For the RG, due to its higher nominal signal, only icy particles that produced a prominent signal were observed. These are therefore compared with the icy particles with the largest volatile content detected by the NG, namely, those calculated with Eq.~\ref{eqn:geom_diameter_volatiles} to have diameters on the order of magnitude of several hundreds of nanometres. The larger diameter values are in agreement with those obtained in Part I with the RG. Considering only particles that have volatiles with comparable equivalent diameters (i.e. larger than 85 nm), the number of icy particles detected by the NG is much higher than that observed by the RG (32464 vs 25). Taking into account that (1) the NG operated for a greater number of days (743 vs 319); (2) the NG had a larger cross section (748 mm$^2$ vs 27 mm$^2$); and (3) the nominal signal of the NG is lower than that of the RG (average density of $4.1\times10^7$ cm$^{-3}$ vs $1.4\times10^8$ cm$^{-3}$), the difference between the two gauges is reduced to a factor of five. This dissimilarity can be explained by the much larger field of view of the NG and by the fact that the signal of the NG is smoother than that of the RG.

Two-thirds of the icy particles detected by the NG have a volatile content with a volume located within the first size bin of Fig.~\ref{fig:size_distr}. This indicates that the NG preferentially detected particles with a small volatile volume, the majority of which were not detectable in the RG data. This can be understood considering that this work relies on an algorithm that could distinguish even small deviations associated with the sublimation of the volatiles of icy particles from the nominal NG signal, whereas in Part I, the higher levels of noise present in the data prevented the effective use of the algorithm and detections were based on manual identification.

\begin{figure}
\centering
\includegraphics[width=\hsize]{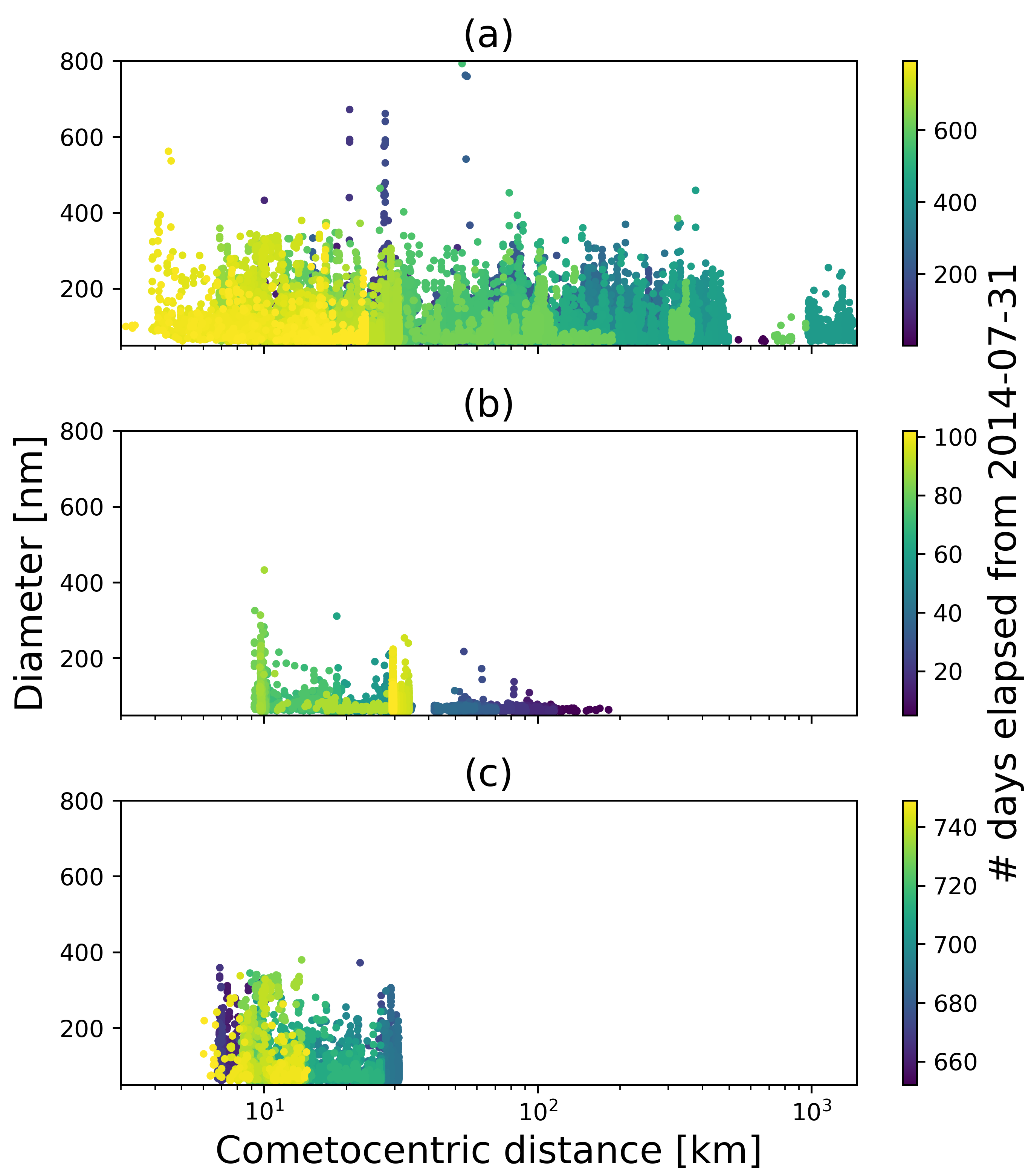}
\caption{Diameters of the equivalent spheres of water ice representing the volatile content of the particle as a function of the cometocentric distance at the time of detection. Diameters are calculated with Eq.~\ref{eqn:geom_diameter_volatiles}. \textit{Panel a}: All icy particles detected over the course of the entire mission (August 2014 to September 2016). \textit{Panels b} and \textit{c}: Icy particles detected at heliocentric distances between 3.0 au and 3.6 au during the inbound trajectory and the outbound trajectory, respectively.}
\label{fig:diameter_to_com_distance}
\end{figure}

The first nine size bins in Fig.~\ref{fig:size_distr} have sufficiently high statistics for a fit (red dashed line) of the form:
\begin{equation}
\label{eqn:fit_size_dist}
\# Events \propto \Bar{d}^{\,\alpha},
\end{equation}
where $\Bar{d}$ is the average diameter of each bin.
The power index $\alpha=-4.79\pm 0.26$ is smaller than those found by other instruments on board Rosetta, such as COSIMA \citep[$\alpha=-0.8\pm 0.1$ for particles larger than 150 $\mu$m and $\alpha=-1.9\pm 0.3$ for particles between 30 $\mu$m and 150 $\mu$m,][]{Merouane_et_al_2016}, GIADA \citep[$\alpha=-1$ for particles smaller than one millimetre,][]{Rotundi_et_al_2015}, and OSIRIS \citep[$\alpha=-3$ for particles smaller than 1 mm][]{Moreno_et_al_2016}. However, this finding is not necessarily a contradiction as they may refer to different combinations of volatiles or refractories. 
The relationship between volatile diameter and cometocentric distance was also investigated (Fig.~\ref{fig:diameter_to_com_distance}). Three periods were explored: (a) the entire mission duration in which the NG was actively acquiring data; (b) the inbound trajectory during which 67P was at a heliocentric distance between 3.6 au and 3.0 au; and (c) the outbound trajectory where 67P was at a heliocentric distance between 3.0 au and 3.6 au. No correlation can be found in any of the three periods considered. This could be due to the complex interplay of parameters (and cometary activity) already proposed in Sect.~\ref{subsec:param_varia_analysis}. Closer to the Sun, sublimation is enhanced. However, larger dust particles may also have been lifted and would have been moving faster due to higher gas activity, and thus reaching Rosetta more quickly \citep[cf. e.g.][]{Tenishev_et_al_2011}. These effects may compensate for each other to some extent.

%%%%%%%%%%%%%%%%%%
%%%%%%%%%%%%%%%%%%
%%%Discussion
%%%%%%%%%%%%%%%%%%
%%%%%%%%%%%%%%%%%%

\section{Discussion}\label{sec:discussion}
All the results presented in this work suggest a clear relationship between cometary activity and the NG detection rate. Additionally, the size distribution of the volatile content of the detected icy particles leads, on the one hand, to values comparable with those derived from the RG and, on the other hand, to a power index higher than those found by GIADA, COSIMA, and OSIRIS. This suggests that for smaller volatile volumes, the index of the size distribution of dust may be different or that volatile content may be size-dependent. Further study is required on this front.

This work provides a proof of concept for the potential use of the NG to analyse the volatile content of icy particles. The results are encouraging, especially when compared with those of other instruments. Nevertheless, there remain issues that need to be addressed. The validity of the results is discussed in this section, as are the complications unique to the type of particles detected.

\subsection{Location of icy particles during sublimation}
\label{subsec:location_during_sublimation}
It is highly unlikely that the detected icy particles had simply passed straight through the NG. This is because the time needed to pass through the sensor, and the duration of the resulting signal, would have been much shorter than the time resolution of the NG itself. Consequently, the detected icy particles must have become stuck to the spacecraft, either to the surface or within the NG or in its vicinity. For simplicity, this work assumes that all detected icy particles were within the NG. 

In this step, we performed a first-order approximation to show that the size distribution power index obtained in Sect.~\ref{subsec:size_distribution} is not affected by the distance between the positions of the icy particles and the NG by considering two identical icy particles. The first icy particle is situated inside the NG and the second is located at a distance $r$ from the NG. The quantities related to these two particles will be identified with the subscripts 1 and 2, respectively. Due to geometric considerations, only a fraction of the sublimated volatiles from the second icy particle will reach the NG. A meaningful scaling function for this would be proportional to $r^{-2}$ and independent of the particle size itself. The relationship between the densities measured by the NG for the first and second icy particles is thus given by $n_2=n_1r^{-2}$. By inserting this relation into Eq.~\ref{eqn:geom_diameter_volatiles}, the resulting relationship between their respective diameters is found to be $d_2=d_1r^{-2/3}$. If all icy particles were at the same distance $r$ from the NG, the values in the histogram in Fig.~\ref{fig:size_distr} would be higher by a factor $r^{-2/3}$. The size distribution power index, however, would remain the same, as the slope is unchanged. This is because the only difference this would make to Eq.~\ref{eqn:fit_size_dist} would be to include the factor of $r^{-2\alpha/3}$, which is independent of the particle diameter.

As the NG signal represents the sum of the contributions from multiple sources at various distances from the NG, the power index calculated in Sect.~\ref{subsec:size_distribution} should remain unaffected.

\subsection{Charge neutrality of icy particles}
\label{subsec:neutrality_agglom}
As explained in Sect.~\ref{subsec:cops_nude_gauge_principle}, only neutral molecules can enter the ionisation region of the NG. This means that any charged icy particles in the immediate vicinity of COPS are not detected (unless they have sufficient kinetic energy to enter the ionisation region, as explained below). It is therefore necessary to understand the process by which particles are lifted from the comet's surface and whether there is any relationship between the neutral and charged icy particles, so as to be able to establish the total amount of particles solely from observations of the neutral component. Any phenomena that could affect the charge of icy particles include: photoelectron emission, interaction with the surrounding plasma, interactions with cosmic rays, and the electric field of the Rosetta spacecraft -- and would further impact the results. 

The assumption thus far that only neutral icy particles can enter the ionisation region is also valid only for low-energy particles. Positively charged particles, having a kinetic energy-to-charge ratio higher than 180 eV/e, can hypothetically enter the NG (180 V is the potential of the inner grid of the NG, cf. Fig.~\ref{fig:nude_gauge}). The same also holds for negatively charged particles having a kinetic energy-to-charge ratio higher than 12 eV/e, which is the potential of the outer grid of the NG (Fig.~\ref{fig:nude_gauge}). As a first approximation to estimate the range of kinetic energies of incoming particles, this study considers spherical icy particles with radii between $10^{-8}\,\mathrm{m}$ and $10^{-6}\,\mathrm{m}$, densities ranging from $800\,\mathrm{kg}$ m$^{-3}$ \citep[GIADA,][]{Fulle_et_al_2016} to $1000\,\mathrm{kg}$ m$^{-3}$ (pure water), and velocities between 0.1 and $10\,\mathrm{m\,s^{-1}}$ \citep[GIADA,][]{Rotundi_et_al_2015,Della_Corte_et_al_2015}. The potential of the icy particles was assumed to be $\sim$4.5 $\mathrm{V}$, as proposed by \citet{Nordheim_et_al_2015}. The equilibrium charge of the icy particles ranges from $\sim$31$\,$e for $10\,\mathrm{nm}$ particles to $\sim$3100 $\mathrm{e}$ for 1$\,\mathrm{\mu}$m particles \citep[][]{Kempf_et_al_2006,Nordheim_et_al_2015}. From the aforementioned considerations, the kinetic energy-to-charge ratios of incoming icy particles were estimated to lie in the range between $3\times10
^{-6}\,\mathrm{eV/e}$ and $4\times10^2\,\mathrm{eV/e}$. It may therefore be possible for some of the more energetic charged particles to enter the NG.

\subsection{Total size of the icy particles}
\label{subsec:dust_size}
The difficulty in providing a reliable estimate of the total volume of volatiles in an icy particle is twofold. Firstly, the results in Fig.~\ref{fig:size_distr} are lower limits, as it is possible that not all of the volatiles entered the ionisation region of the NG during the measurements. Features that were spread over multiple measurements were also truncated and only the data point with the highest density was retained. Secondly, in the case of volatiles enclosed within refractories, it is not clear what fraction of the volatiles is able to escape or escapes slowly and below our detection limit. 
It is also not possible to determine the proportion of refractories to volatiles in the icy particles that are detected, or whether or not they even contain any refractories to begin with. Any attempt to estimate the full size of an icy particle would therefore prove futile.

\subsection{Origin of volatiles}
\label{subsec:provenance}
The volatiles measured in this study originate from icy particles that are sublimated inside COPS. This does not in any way constrain the initial mass that the particle may have had when it was first ejected from the surface of 67P. Even if the icy particles detected came directly and structurally unaltered from 67P, they could still have originated from a larger particle that fragmented on its way to the spacecraft and lost an undetermined portion of its volatiles (depending on the trajectory, illumination conditions, initial size, composition, and porosity). Possible scenarios for the fragmentation of a larger particle into smaller ones include impacts on the orbiter itself, charging processes of the icy particles in combination with spacecraft charging \citep[][]{Hill_1981}, or the evaporation of semi-volatiles, organics, and salts that were acting as a glue in the original particle \citep[][]{Komle_et_al_1996}. 

It is thus entirely possible that not all events detected by the NG represent independent icy particles, as multiple fragments originating from a larger parent may have been detected. However, the effect of fragmentation does not appear to be significant \citep[][]{Gerig_et_al_2018}. Conversely, a jump in density may not have been generated by a single particle, but several particles, in fact, impacting the NG nearly simultaneously \citep{Fulle_2015}.

%%%%%%%%%%%%%%%%%%
%%%%%%%%%%%%%%%%%%
%%%Conclusion
%%%%%%%%%%%%%%%%%%
%%%%%%%%%%%%%%%%%%

\section{Conclusions}\label{sec:conclusion}
In this work, we carry out a first study on the detection of the volatile content of icy particles originating from 67P using the NG of COPS.
Based on the observed density features, $\sim$67000 icy particles were found. The feasibility of the approach used was demonstrated: although the NG was designed for other purposes, it nevertheless allows for the study of the volatile content of icy particles. So far, no other instrument in the Rosetta mission has been employed to perform such an extensive analysis of the volatile content of dust.

We performed following statistical analyses of the icy particles detected: (i) a study of the influence of the Rosetta spacecraft's location on the number of detections and NG detection rate; (ii) an investigation into how the NG detection rate varied during the mission; (iii) a comparison of the NG detection rate with other quantities related to cometary activity; and (iv) an estimation of the minimum quantity of water ice present in these icy particles. 

The complex interplay of parameters such as the nominal NG signal, the off-nadir angle, the sub-spacecraft latitude, the phase angle, the cometocentric distance, and the heliocentric distance was studied by investigating the effects of each variable separately. Dependencies between some of these variables and the NG detection rate were found.

Moreover, there were many more icy particles detected during the inbound orbit than during the outbound orbit. This result is consistent with reports of 67P shedding a dust mantle during its inbound trajectory. This process may therefore be extended also to the volatile content of icy particles and not only to refractories \citep[][]{Guilbert_Lepoutre_et_al_2014,Schulz_et_al_2015}.  

The results were also found to be in accordance with those observed in other studies, such as \citet{Snodgrass_et_al_2016}, \citet{Hansen_et_al_2016}, and Part I. The strong correlation between the various datasets proves that the NG icy particle detection rate is related to cometary activity.

Finally, the diameters of water ice spheres equivalent to the volatile content of the icy particles were calculated and a size distribution with a power index of $\alpha=-4.79\pm 0.26$ was found. Further studies will be required to validate this value, as only volatiles between 60 and 390 nanometres in equivalent diameter were included in its calculation.

Progress on this front will require a multi-disciplinary collaboration between various experts in the 67P scientific community, combining comet simulation with experiments to reproduce the results of space-based measurements in the laboratory.

%%%%%%%%%%%%%%%%%%
%%%%%%%%%%%%%%%%%%
%%%Acknowledgements
%%%%%%%%%%%%%%%%%%
%%%%%%%%%%%%%%%%%%

\begin{acknowledgements}
Work at the University of Bern was funded by the State of Bern, the Swiss National Science Foundation (200020\textunderscore182418), and the European Space Agency through the Rosetta data fusion: Dust and gas coma modelling (5001018690) grant. We gratefully acknowledge the many engineers, technicians, and scientists involved in the Rosetta mission, and in particular in the ROSINA instrument. Rosetta is an European Space Agency (ESA) mission with contributions from its member states and NASA. All ROSINA flight data have been released to the PSA archive of ESA and to the PDS archive of NASA.

We would like to show our gratitude to the International Space Science Institute (ISSI) team ``Characterization Of Cometary Activity Of 67P/Churyumov-Gerasimenko Comet''
for their comments and valuable inputs.

The author would also like to thank Carsten G{\"u}ttler for
his detailed suggestions which helped us to considerably improve this article.
\end{acknowledgements}

\bibliographystyle{aa} % style aa.bst
\bibliography{volatiles_detection_with_ROSINA_COPS_Part_II_nude_gauge} % the .bib file

\end{document}